\documentclass{article}
\usepackage[utf8]{inputenc}

\usepackage{xcolor}
\usepackage{amsfonts,amsmath,amssymb,amsthm}
\usepackage[colorlinks = true,
            linkcolor = blue,
            urlcolor  = blue,
            citecolor = blue,
            anchorcolor = blue]{hyperref}
\usepackage{soul}
\usepackage{bm}
\usepackage{graphicx}
\usepackage{tikz}
\usepackage{pgfplots}
\usepackage{authblk}
\usepackage{url}
\newcommand{\wh}{\widehat{W}}

\newtheorem{remark}{Remark}

\providecommand{\keywords}[1]
{
  \small	
  \textbf{\textit{Keywords---}} #1
}

\title{Smoothening block rewards: How much should miners pay for mining pools?}
\author[1,2]{Axel Cortes-Cubero\footnote{Main author} [\url{axel.cortescubero@protocol.ai}]}
\author[1,2]{\\Juan P. Madrigal-Cianci [\url{juan.madrigalcianci@protocol.ai}]}
\author[1,2]{\\Kiran Karra [\url{kiran.karra@protocol.ai}]}
\author[1,2]{\\Zixuan Zhang [\url{zixuan@protocol.ai}]}
\affil[1]{CryptoEconLab\footnote{\url{https://cryptoeconlab.io}}}
\affil[2]{Protocol Labs\footnote{\url{https://protocol.ai}}}

\date{\today}

\begin{document}

\maketitle


\begin{abstract}
    The rewards a blockchain miner earns vary with time. Most of the time is spent mining without receiving any rewards, and only occasionally the miner wins a block and earns a reward. Mining pools smoothen the stochastic flow of rewards, and in the ideal case, provide a steady flow of rewards over time. Smooth block rewards allow miners to choose an optimal mining power growth strategy that will result in a higher reward yield for a given investment. We quantify the economic advantage for a given miner of having smooth rewards, and use this to define a maximum percentage of rewards that a miner should be willing to pay for the mining pool services.
\end{abstract} \hspace{10pt}





\newpage

\section{Introduction}
Miners in a blockchain receive block rewards as an incentive for participating in distributed consensus. Miners must compete to win the right to add a block to the blockchain, and rewards are distributed to the miner who adds the block to the chain for a given time-step. Different blockchain protocols vary in the way they distribute these rewards.  For example, in Bitcoin~\cite{nakamoto2008bitcoin}, a miner wins the right to add a block to the blockchain (i.e. mine a block) by solving a complex mathematical problem. The difficulty of this problem is adjusted such that blocks are mined \textit{approximately} every ten minutes. A miner's probability of mining a block is proportional to that miner's share of the total network hashing capacity. A similar situation exists in Filecoin \cite{filecoin_whitepaper}, where miners who provide storage have the opportunity to mine blocks. In Filecoin, a random number $N$ of blocks (distributed according to a Poisson distribution with a mean of 5 blocks per epoch) are added to the chain every 30 seconds. The probability of mining a block is proportional to the miner's (adjusted) storage power relative to the network's (adjusted) storage power. In both of these scenarios (stochastic arrival of rewards), the time between \textit{receiving} rewards\footnote{Not to be confused with the time it takes to add a block by any miner.} is highly variable for an individual miner. 

Conversely, protocols such as Ethereum~\cite{wood2014ethereum} define the number of blocks that are to arrive in a given time-period $T_b$ in a deterministic way. This can be modeled in a probabilistic framework by setting the distribution that determines the time between blocks, $t_n$, to be a random variable sampled from a \textit{generalized distribution that has a unit mass at $T_b$}, i.e., $t_n\sim\delta_{T_b}$. In both stochastic and deterministic arrival of rewards cases, the lottery process that determines who gets to mine a block remains.

Participating in a \textit{mining pool} is one way to reduce the variability in rewards arrival times. A mining pool is a group of cryptocurrency miners who combine their computational resources to gain more power, thus increasing their probability of winning rewards (i.e. mining blocks). The rewards from mining blocks are then distributed to participants of the pool through some mechanism, after fees for participation are charged. 

In this paper, we focus on the stochastic arrival of rewards and derive a probability distribution describing the rewards a miner can expect to receive in a given time window. We show that stochastic rewards converge to smooth rewards as the considered time-window is expanded (c.f. Equation \eqref{eq:expected_reward}). From the distribution, we also derive the expected waiting time before a miner wins rewards. We use this distribution to determine how much miners should pay for entering mining pools, with the goal of maximizing the returns from participation. 

Next, we apply these results to decisions a miner would take when participating in a mining pool. Standard decision theory dictates that miners should make decisions that maximize expected returns. However, the St. Petersburg paradox~\cite{stpetersburg_paradox} indicates that expected returns will not be the same as time-averaged returns. It is evident that individual entities (whether they be miners or mining pools) are more interested in maximizing their time-averaged returns rather than expected returns. We apply recent insights from Peters et al. ~\cite{peters2011time} and derive several results pertaining to miner strategy in the context of using mining pools. 
The primary contributions of this work are:
\begin{itemize}
    \item We derive the probability distribution to predict a miner's rewards within a given time frame, and compute the average waiting period that it takes a miner to receive their reward.
    \item We show that stochastic rewards evolve into smoother, consistent rewards as the observation window expands.
    \item We derive a formula for estimating the growth of a miner's wealth, for a given investment strategy, which is optimal under \textit{time-averaged returns} (c.f. Sections \ref{sec: St Petes} and \ref{sec: miner wealth}).
    \item We explore decision-making strategies for miners, contrasting \textit{expected returns} with \textit{time-averaged} returns (c.f. Section \ref{sec: miner wealth}).
\end{itemize}

The rest of this work is organized as follows. In Section \ref{sec: prev work} we present a brief review of prior research related to mining pools. Section \ref{sec:bitcoinpois} describes how block arrivals can be modeled as a Poisson process (although it can readily be generalized to other, arbitrary counting processes). We derive the probability distribution for miner rewards in Section \ref{sec:prob_dist}. In Section \ref{sec: St Petes} we recall the \textit{St. Petersburg paradox}, and present a solution to it based on arguments from \textit{ergodicity economics} \cite{peters2011time}. In that section we also introduce a \textit{Time-Averaged, Non-Ergodic (TANE) utility function}, which is central to the results in \ref{sec: miner wealth}. In that section, we study the wealth and decision-making  process of the miners under a TANE utility function. We derive some results related to the optimal allocation of a miner's wealth under this utility function, as well as deriving a maximal price that a miner should pay to belong to a pool.  Lastly, we present our conclusions in Section \ref{sec:concl}.

\section{Previous Work} \label{sec: prev work}

Prior work in mining pools can be broadly categorized as either:
\begin{enumerate}
\item \textit{Mining pool selection,} which is concerned with understanding how should a miner choose which pool(s) to participate in and,
\item \textit{adversarial aspects of mining pools,} which is concerned with studying how can miners \textit{game} the pool (via miner collusion, for example), with the aim of obtaining, e.g., a larger amount of rewards compared to their honest peers.
\end{enumerate}

Regarding \textit{mining pool selection}, Qin et al.~\cite{qin2018research} investigate various selection strategies for blockchain mining pools by modeling the problem as a risk-based decision, since each mining pool can have a different risk/reward tradeoff. Chatzigiannis et al.~\cite{chatzigiannis2022diversification} address the same problem of mining pool selection, but approach the optimal solution from an optimization perspective. Liu et al.~\cite{liu2018evolutionary} also address mining pool selection but from an evolutionary game perspective. 

Numerous works have also explored \textit{adversarial aspects of mining pools}. Some adversarial research focuses on gaining extra rewards from participation, whether it be through group strategies~\cite{liu2017intelligent}, or strategies for individuals participating to decide amongst various available pools~\cite{wang2019pool}. Other tracks of adversarial research are involved in game theoretic formulations of how miners can act within a pool, for example whether to share a solution with the pool if one is found and related incentive mechanisms in that space~\cite{qin2020optimal, schrijvers2017incentive, laszka2015bitcoin}. The common thread among these works is that they are focused on ways that the mining pool construct can be exploited to gain ``alpha'' in a rewards context. 

The first approach is more akin to that of a \textit{resource allocation} problem, while the second one is more concerned with the \textit{game-theoretical} aspects of participation. 

Our work builds upon but differs from these two approaches in the following way: we are concerned with optimal miner strategy when a miner has limited resources to invest into a single pool. We don't take an adversarial perspective, but focus on an optimal strategy for a miner to determine when to invest more resources into a pool, versus holding enough reserve to account for the stochastic arrival of rewards that are a result of the mining mechanisms in place in protocols such as Bitcoin and Filecoin. 

\section{Modeling Bitcoin block arrival as a Poisson process} \label{sec:bitcoinpois}
Block arrival times in Bitcoin can be modeled as a Poisson process \cite{lewenberg2015bitcoin, rosenfeld2011analysis, kroese2013handbook}. The level of difficulty in the Proof of Work (PoW) protocol  is adjusted such that a block is expected to be added approximately every 10 minutes. It is however, not exactly known when a block will arrive, but in an interval of $T$ minutes, $E=T/10$ blocks will be added to the chain per minute in expectation.

We can model this by considering the number of blocks $w\in\mathbb{N}_{+}$, to arrive within a time interval $T\in\mathbb{R}_{++}$
to be drawn from a Poisson distribution, with mean $E$,

\begin{eqnarray}
    w\sim {\rm Poisson}[E].
\end{eqnarray}
We note that while in Bitcoin, the distribution above is induced by the PoW mechanics, other protocols such as Filecoin employ an Expected Consensus mechanism \cite{filecoin_whitepaper}, where such a stochastic, Poisson-distributed number of blocks is explicitly written into the protocol. 

While the Poisson process induced above can accurately describe PoW and Expected Consensus protocols, like Bitcoin and Filecoin, respectively, in most Proof of Stake (PoS) protocols, such as Ethereum, the number of Blocks that arrive in a given time period is deterministic and fixed (more accurately, the maximum number of blocks is fixed, but there could be fewer blocks if some of the blocks fail to be added to the chain). The rest of this paper will use the PoS/Expected consensus model, described by the Poisson process, but all the results could be easily modified to apply to PoS protocols by replacing this Poisson process by a deterministic one.

\section{Deriving miner reward probability distribution}\label{sec:prob_dist}

We can  quantify how likely an individual miner is to win a block in a given time interval. Suppose the $i$-th miner controls an amount $p_i\in \mathbb{R}_+$ of consensus power, with total network power given by $P\in \mathbb{R}_+$ (For PoW this corresponds to total computational power). The miner $i$ has a probability
\begin{eqnarray}
    q_i=\frac{p_i}{P},
\end{eqnarray}
of being the winner of each block. If in a given period, $T$ there were $w$ blocks added to the chain, the probability that $v\in\mathbb{N}_+,$ $v\leq w$, of those $w$ blocks will be won by the $i$-th miner is given by a Binomial distribution, with a probability density function (PDF) given by
\begin{eqnarray}
    f_{\rm Binomial}\left(v,w,\frac{p_i}{P}\right)=\frac{w!}{v!(w-v)!}\left(\frac{p_i}{P}\right)^v\left(1-\frac{p_i}{P}\right)^{w-v}.
\end{eqnarray}

We now derive the distribution for the total rewards obtained by a miner $i$ in a time interval $T:=[n,n+N]$. Given some expected block arrival rate $E\in\mathbb{R}_+,$ let $\{p^i_k\}_{k=n}^{n+N}$ and $\{P_k\}_{k=n}^{n+N}$ denote a sequence of miner and network powers, respectively, between epochs $n$ and $n+N$, and similarly, let $\{M_k\}_{k=n}^N,\ M_k\in\mathbb{R}_+\ \forall k\in\mathbb{N},$ denote a sequence of block rewards produced by the network in the same interval\footnote{Note that this formulation allows the flexibility of considering $\{M_k\}_{k=n}^N$, $\{p^i_k\}_{k=n}^N$, $\{P_k\}_{k=n}^N$ as random variables driven by some known stochastic process (which can account for varying rewards such as when taking \textit{miner tips} into account)}. Let us assume that \begin{enumerate}
    \item[(i)] \textit{Each sequence is made out of independent components,} and
    \item[(ii)] \textit{The sequences $\{M_k\}_{k=n}^N$, $\{p^i_k\}_{k=n}^N$, $\{P_k\}_{k=n}^N$ are independent of each other.}
\end{enumerate} Furthermore, let $\mathcal{M}^i(n,N)\geq0$ denote the total rewards obtained by the $i^\text{th}$ miner between epochs $n$ and $n+N$. From this, we can derive the PDF of $\mathcal{M}^i(n,N)$, denoted by
\begin{eqnarray}  
f(\mathcal{M}^i(n,N);\{M_k\}_{k=n}^N,\{p^i_k\}_{k=n}^{n+N},\{P_k\}_{k=n}^{n+N},E)
\end{eqnarray}

We start by computing $f_1(v_n^i;p_n^i,P_n,E)$, which is the PDF associated to the event that in the time period $n$, the $i$-th miner will get to write $v_n^i$ blocks. This can be computed by marginalizing the Poisson distribution for the total number of blocks to be written, with respect to the binomial distribution which fixes how many of these blocks will be won by the $i$-th miner,
\begin{eqnarray}
f_1(v_n^i;p_n^i,P_n,E)&=&\sum_{w_n=0}^\infty f_{\rm Binomial}\left(v_n^i,w_n,\frac{p_n^i}{P_n}\right)f_{\rm Poisson}(w_n,E) \label{eq:blw} \\
&=&\sum_{w_n=0}^\infty \frac{E^{w_n}e^{-E}}{v_n^i!(w_n-v_n^i)!}\left(\frac{p_n^i}{P_n
}\right)^{v_n^i}\left(1-\frac{p_n^i}{P_n}\right)^{w_n-v_n^i}.
\end{eqnarray}

From this, we can compute the PDF $f_2(m_n^i;M_n,p_n^i,P_n,E)$, that in the time period $n$ the $i$-th miner earns an amount $m_n^i$ of reward, which is given by
\begin{eqnarray}
    f_2(m_n^i;M_n,p_n^i,P_n,E)=\sum_{v_n^i=0}^{\infty}\delta({m_n^i}-M_n v_{n}^i)f_1(v_n^i;p_n^i,P_n,E),
\end{eqnarray}
where $\delta(x)$ is a Dirac delta function.

Under assumptions (i) and (ii) above, summing this distribution over several time periods yields

\begin{eqnarray}
    &&f(\mathcal{M}^i(n,N);\{M_k\}_{k=n}^N,\{p^i_k\}_{k=n}^{n+N},\{P_k\}_{k=n}^{n+N},E)\nonumber\\
    &&\,\,\,\,\,\,\,=\delta\left(\mathcal{M}^i(n,N)-\sum_{k=0}^N m_{n+k}\right)\nonumber\\
    &&\,\,\,\,\,\times\int_{0}^\infty\dots\int_0^\infty \left[\prod_{k=0}^Nf_2(m_{n+k}^i;M_{n+k},p_{n+k}^i,P_{n+k},E) \right] \underbrace{\mathrm{d}m_n\dots,\mathrm{d}m_{n+N}}_\text{$:=\mathrm{d}\mathcal{M}$}.
\end{eqnarray}




\subsection{Expected reward}

Using this probability distribution, we can now compute the expected value of the reward the $i$-th miner will obtain within the time periods from $n$ to $n+N$,
\begin{eqnarray}
    \langle\mathcal{M}^i(n,N)\rangle&=&\int_0^\infty \,\mathcal{M}f(\mathcal{M}^i(n,N);\{M_k\}_{k=n}^N,\{p^i_k\}_{k=n}^{n+N},\{P_k\}_{k=n}^{n+N},E)\mathrm{d}\mathcal{M}\nonumber\\
    &=& E\sum_{k=0}^N M_{n+k}\frac{p_{n+k}^i}{p_{n+k}}. \label{eq:expected_reward}
\end{eqnarray}

Eq.~\ref{eq:expected_reward} indicates that the expected reward for stochastically arriving rewards behaves the same as smoothly arriving rewards. That is, every miner receives their corresponding portion of total rewards.

\subsection{Variance of reward}

The first key difference between smooth rewards vs stochastic block rewards is the variance in reward arrival times. While smooth rewards have zero variability over time (each miner knows exactly how much reward to expect at each time period), stochastic block rewards have variance:

\begin{eqnarray}
    \text{Var}(\mathcal{M}^i(n,N))=\langle\mathcal{M}^i(n,N)^2\rangle-\langle \mathcal{M}^i(n,N)\rangle^2
\end{eqnarray}
Furthermore, it can be shown that

\begin{eqnarray}
\text{Var}(\mathcal{M}^i(n,N))=e^{-E}\sum_{m=0}^NE^2M_{n+m}^2\left[1+\frac{p_{n+m}^i}{P_{n+m}}\left(1-\frac{p_{n+m}^i}{P_{n+m}}\right)\left(\text{Ei}(E)-\log(E)-\Gamma\right)\right]
\end{eqnarray}
where $\text{Ei}(x)$ is the exponential integral function, and $\Gamma$ is the Euler-Mascheroni constant.

\subsection{Expected time before reward}

An important difference between stochastic block rewards and smooth rewards, is that with stochastic rewards, if the total network power becomes sufficiently larger than the $i$-th miner's power, there is a chance that mining is not viable (a miner would run out of resources to keep mining, before receiving any reward). Therefore, it is important to estimate the amount of time a miner should expect to wait before receiving their first reward.  

If total reward and power remain constant, the probability of winning $v_n^i$  blocks over a period of $x$ epochs is $f_1(v_n^i,p_n^i,P_n,xE)$, where $f_1$ is defined in Eq.~\ref{eq:blw}. For simplicity, we will assume for the reminder of this work, that this is indeed the case \footnote{The case of non-constant power, while more general, is left for future work}. Intuitively, this assumption should hold reasonably well in small to medium time horizons; as it seems unlikely that power will have a large, sudden change.

The cumulative distribution of number of time periods, $x\in\mathbb{N}_+$, the miner has to wait to win their first block is then given by
\begin{eqnarray}
    F_4(x,p_m^i,P_n,E)&=&1-f_1(0,p_n^i,P_n,xE)\nonumber\\
    &=& 1-\sum_{w_n=0}^\infty \frac{(xE)^{w_n}e^{-xE}}{w_n!}\left(1-\frac{p_n^i}{P_n}\right)^{w_n}\nonumber\\
    &=&1-\frac{e^{-xE}}{e^{-\left[(xE)\left(1-\frac{p_n^i}{P_n}\right)\right]}}\sum_{w_n=0}^\infty\left[(xE)\left(1-\frac{p_n^i}{P_n}\right)\right]^{w_n}e^{-\left[(xE)\left(1-\frac{p_n^i}{P_n}\right)\right]}\frac{1}{w_n!}\nonumber\\
    &=&1-e^{-xE\frac{p_n^i}{P_n}},
\end{eqnarray}
which is the Cumulative Distribution Function (CDF) of an exponential distribution with parameter $E\frac{p_n^i}{P_n}$. The 
 PDF modeling waiting times between winning blocks is then given by
\begin{eqnarray}
    f_4(x,p_m^i,P_n,E)=\frac{\mathrm{d}F_4(x,p_m^i,P_n,E)}{\mathrm{d}x}=E\frac{p_n^i}{P_n}e^{-xE\frac{p_n^i}{P_n}}.
\end{eqnarray}

\begin{remark}
$F_4$ is an exponential distribution, implying that the block-winning process is \textit{memoryless}. That is, if $t_1,t_2$ denote the waiting times between any two rewards, $t_1$ and $t_2$ are statistically independent of each other.
\end{remark}

\begin{remark} The distribution for these waiting times is a well-known property of Poisson processes (see, e.g., \cite[p. 172]{kroese2013handbook}), which in our case follows as a consequence of having the block-arrival process follow a Poisson distribution.
\end{remark}

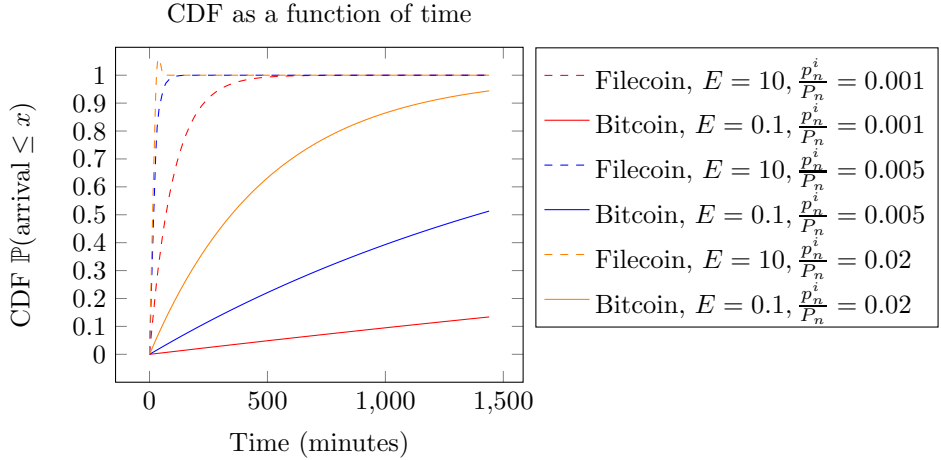
\begin{figure}[h] 
\label{fig:cdf}
\centering
\begin{tikzpicture}
\begin{axis}[
    width=7cm,
    xlabel = Time (minutes),
    ylabel = {CDF $\mathbb{P}(\text{arrival}\leq x)$},
    title= CDF as a function of time,
    ytick={0,0.1,0.2,0.3,0.4,0.5,0.6,0.7,0.8,0.9,1
    },
            legend pos=outer north east,
        legend cell align=left,
        smooth,
]
\addplot [
    domain=-0:60*24, 
    samples=50, 
    color=red,
    style=dashed
]
{1-exp(-10*0.001*x};
\addlegendentry{Filecoin, $E=10, \frac{p_n^i}{P_n}=0.001$}
\addplot [
    domain=-0:60*24, 
    samples=50, 
    color=red
]
{1-exp(-0.10*0.001*x};
\addlegendentry{Bitcoin, $E=0.1, \frac{p_n^i}{P_n}=0.001$}

\addplot [
    domain=-0:60*24, 
    samples=50, 
    color=blue,
    style=dashed
]
{1-exp(-10*0.005*x};
\addlegendentry{Filecoin, $E=10, \frac{p_n^i}{P_n}=0.005$}
\addplot [
    domain=-0:60*24, 
    samples=50, 
    color=blue
]
{1-exp(-0.10*0.005*x};
\addlegendentry{Bitcoin, $E=0.1, \frac{p_n^i}{P_n}=0.005$}

\addplot [
    domain=-0:60*24, 
    samples=50, 
    color=orange,
    style=dashed
]
{1-exp(-10*0.02*x};
\addlegendentry{Filecoin, $E=10, \frac{p_n^i}{P_n}=0.02$}
\addplot [
    domain=-0:60*24, 
    samples=50, 
    color=orange
]
{1-exp(-0.10*0.02*x};
\addlegendentry{Bitcoin, $E=0.1, \frac{p_n^i}{P_n}=0.02$}

\end{axis}
\end{tikzpicture}
\caption{Visualization of the CDF of expected rewards for various parametrizations of individual miner power ratios}
\end{figure}

We visualize this distribution for various power ratios for Bitcoin ($E$=0.1) and Filecoin ($E$=10) in Fig.~\ref{fig:cdf}.

The expected number of time periods a miner needs to wait before winning their first block is then given by 
\begin{eqnarray}
    \langle x\rangle=\int_0^\infty x\,f_4(x,p_m^i,P_n,E)\,\mathrm{d}x=\frac{P_n}{Ep_n^i}.
\end{eqnarray}
Notice that this waiting time could be made shorter, either by increasing the miner's proportion of the network power, or by increasing the number of expected blocks for each time period (by lowering the difficulty level in Bitcoin, or in Filecoin by changing the parameter $E$ used in the expected consensus algorithm).

Additionally,
\begin{eqnarray}
    \text{Var}(x)=\int_0^\infty \left(x^2-\langle x\rangle^2\right)f_4(x,p_m^i,P_n,E)\,\mathrm{d}x=\left(\frac{P_n}{Ep_n^i}\right)^2.
\end{eqnarray}
From this one can also compute lower bound estimates on the bankruptcy probability of going bankrupt due to lack of waiting too long before receiving rewards. Given an initial wealth $W^i_0$, and a maintenance cost of $C^i$ per epoch, in the absence of earnings, a miner will go bankrupt in $x^{*,i}=\left\lceil W^i_0/C^i\right\rceil$ epochs (where $\left\lceil \cdot\right\rceil$ is the \textit{ceiling} operator). Thus, the probability of going bankrupt due to lack of winning (i.e., waiting a time $x>x^{*,i}$) is given by:
\begin{align}
    \mathbb{P}(x\geq x^{*,i})&=1-\mathbb{P}(x\leq x^{*,i})\\&=1-\int_0^{x^{*,i}} f_4(x,p_m^i,P_n,E)\,\mathrm{d}x
\end{align}

\section{The St. Petersburg paradox, time averaging vs ensemble averaging} \label{sec: St Petes}


Equation \eqref{eq:expected_reward} established that the expected miner reward for a period of time is the same under the smooth and lottery based reward. It is however wrong to assume that the expected reward is the only quantity relevant to the miner finances, and their decision to mine. This point is exemplified in the St. Petersburg paradox \cite{peters2011time}. The St. Petersburg Paradox is a classical problem in probability and decision theory. It was introduced by Daniel Bernoulli in a paper to the St. Petersburg Academy in 1738 \cite{bernoulli22originally}. The paradox highlights a situation where traditional expected monetary value theory seems to fail to provide a reasonable decision guideline. The paradox arises from the following game:\\

\textbf{St. Petersburg Paradox I. The game}\textit{ You are invited to play a game where a fair coin will be flipped repeatedly until it comes up heads. If the first heads appears on the $n^\text{th}$ flip, you receive $2^{n-1}$ dollars. The question is: How much should you be willing to pay to play this game?}\\

According to \textit{classical decision theory,} a rational person would only play the game if their expected profit is larger than the cost to play. Mathematically, one can compute the expected monetary value (EMV) of the game  by multiplying the payout of each possible outcome by its probability and then summing these products. It is easy to show that the EMV derived from this game is given by \begin{align}
    \mathbb{E}[\mathsf{Utility}]=\sum_{n=0}^\infty 2^n \mathbb{P}(\text{win $n$ times in a row})=\sum_{n=0}^\infty2^n\times \left(\frac{1}{2^n}\right)=\infty,
\end{align}
Which means that the game has \textit{infinite} expected utility.\\ 

\textbf{St. Petersburg Paradox II. The paradox} \textit{Since the game above has an infinite expected payoff, any rational person would be willing to pay any finite amount of money to play the game (according to classical decision theory). However, in reality, most people would not be willing to pay very much at all to play the game, let alone an ``infinite" amount of money. This discrepancy between the theoretical expected value and what people are actually willing to pay is at the heart of the paradox.}\\

There are several solutions to the St. Petersburg paradox \cite{bernoulli22originally,peters2011time}, which rely upon techniques such as considering a logarithmic utility profile as a function of the wealth, $W$ (i.e., setting $\mathsf{Utility}(W):=\log(W)$ rather than $\mathsf{Utility}(W):=W$, as is done above) \cite{bernoulli22originally}, or by utilizing arguments from   so-called \textit{ergodicity economics} \cite{peters2011time} to argue that the player of a game does not care about the \textit{ensemble average} (i.e., the average outcome of the experiment if it where to be repeated in an infinite number of parallel universes), but rather the long-time average of their specific game \cite{peters2011time}, which are different if the system is not ergodic\footnote{Given $N$ independent stochastic processes $\{X_t^{(i)}, t\geq 0\}, i=1,2,\dots,N $ driven by the same underlying dynamics or physical phenomena, we say that the process $X$ is \textit{ergodic} if it holds that $$
\lim_{N\to\infty}\frac{1}{N} \sum_{i=1}^N X^{(i)}=\lim_{T\to\infty}\frac{1}{T}\int_{0}^\infty X^{(i)}\mathrm{d}t, \ \forall i\in\mathbb{N},$$ i.e., if the ensemble average of the process is the same as the long-time average.}. While these two approaches result in the exact same mathematical expression for the expected utility, we focus on the approach presented in \cite{peters2011time}. In this setting, the authors define the time-averaged wealth $\wh(T)=\wh_T$ of a player (miner in our context) up-until time $T$ as 
\begin{align}
    \wh_T=W_0e^{g T}, \label{eq:aw}
\end{align}
with $g\in\mathbb{R}$ the so-called \textit{average growth rate}, defined as 
\begin{align}
g:=\sum_{k\in\mathbb{N}}q_k\log\left(\frac{W_0-c_k+\mathcal{M}_k}{W_0}\right), \label{eq:agr}
\end{align}
with $q_k$ the probability of winning the game $k\in\mathbb{N}$ times, and $c_k$, $\mathcal{M}_k$ the cost of playing  and  the expected earnings at round $k$, respectively. 
Thus, in their setting, Peters et al. argue that playing a game is only rational whenever $g>0$; otherwise it is easy to see that the player will go bankrupt (in expectation). Said differently, a player should not participate in the game if the cost of playing is larger that $c^*\in\mathbb{R}_+$, with $c^*$ such that  $g(c^*)=\sum_{k\in\mathbb{N}}q_k\log\left(\frac{W_0-c^+\mathcal{M}_k}{W_0}\right)=0$.

One key insight of \cite{peters2011time} is that this quantity  depends on the initial wealth of the player. A player who starts with a larger amount of wealth should be willing to pay a higher price to pay the game, as their wealth will grow faster with higher initial wealth.  By contrast, a player who doesn't have enough wealth should not play the game, unless they can pay a lower price, as their wealth is expected to be reduced. 

Notice that it follows from Jensen's inequality that $g$ can be upper bounded by a function $\tilde g$ given by 

\begin{align}
\tilde{g}&:=\log\left(\sum_{k\in\mathbb{N}}q_k \frac{W_0-c_k+\mathcal{M}_k}{W_0} \right)\\
&=\log\left(\mathbb{E}_{q}\left[\frac{W_0-c+\mathcal{M}}{W_0}\right]\right)\\
&=\log\left(1+W_0^{-1}\mathbb{E}_{q}\left[\mathcal{M}-c\right]\right), \label{eq:ub}
\end{align}
where $\mathbb{E}_{q}$ denotes the expectation induced by the probability mass function of $q_n$.  Equation ~\eqref{eq:ub} implies an upper bound on the profitability condition; if $\tilde{g}<0$, so will $g$, and this occurs whenever $\mathbb{E}_{q}\left[c\right]>\mathbb{E}_{q}[\mathcal{M}]$. 

Next, we utilize  Equations \eqref{eq:aw} and \eqref{eq:agr} to 
model the wealth process of a miner and discuss their utility and ideal behaviour under this notion of this Time-Averaged, Non-Ergodic (TANE) utility. In this setting, $c_k$ represents the Operational Expenses (OpEx) of the miner (including but not limited to, hardware, maintenance, electricity, human capital, etc) at epoch $k$, and similarly $\mathcal{M}_k$ represents the rewards on that epoch.

\section{Miner behaviour under TANE  utility }\label{sec: miner wealth}
We now apply the results of \cite{peters2011time} to the problem of miner return on investment, starting with the case of the stochastic block reward.

\subsection{Wealth growth rate from random block reward}

Suppose a miner starts with an amount of wealth/investment $W>0$. They must first arrange their budget, and decide how much to spend on Capital Expenditures (CapEx) to enable onboarding new capacity (buying  equipment) and how much budget to leave so they are able to run their equipment for a given amount of time before making any profit.

Suppose the miner splits their investment in $\gamma W$ for acquiring equipment, and $(1-\gamma)W$ for running their equipment, where $\gamma\in(0,1)$. 

The total amount of consensus power the miner will obtain is given by
\begin{eqnarray}
    p^i=C_e(\gamma W),
\end{eqnarray}
where $C_e(x)$ is a price function that models how much equipment we can purchase for an amount of money $x$. Generally one could expect that $C_e(x)$ is sublinear to account for effects of economies of scale.

The miner will be able to run this equipment for an amount of time $t$ at a cost $C_r(t,C_e(\gamma W))$, where $C_r(t,p)$ is a function that models the cost to run $p$ amount of equipment for time $t$. The maximum amount of time the miner can run their equipment without going bankrupt, $T_{\rm max}$, is fixed by
\begin{eqnarray}
C_r(T_{\rm max},C_e(\gamma W))=(1-\gamma)W 
\end{eqnarray}

We make the {\it simplifying assumption} that costs grow linearly with amount of onboarded capacity and running time, such that

\begin{align}
C_e(\gamma W)=\gamma Wc_e,\,\,\,\,\,C_r(t,C_e(\gamma W))=t\gamma Wc_ec_r, \label{eq: lin_cond}
\end{align}
for some constant rates $c_e$ and $c_r$. The maximum amount of time a miner can run their equipment is then,
\begin{equation}
    T_{\rm max}=\frac{1-\gamma}{\gamma c_ec_r}. \label{eq:tmax}
\end{equation}
Put differently, a miner will go bankrupt if they do not win at least once in a time $T_{\rm max}$. Note the subtlety that it is not enough to win before $T_{\max}$ elapses, as the wealth process is not memoryless. The win needs to at least account for the expenses in that time, otherwise, if we want \eqref{eq: lin_cond}, \eqref{eq:tmax} to hold, the miner would need to immediately scale up-or-down their operations (i.e., buy and sell equipment) after each round.

As previously shown, the amount of time the miner must wait before receiving a reward follows an exponential distribution. Suppose the total network power before the miner joined was $P_0$. The total network power, including the new miner is now $P=P_0+\gamma W c_e$. The probability of  waiting a time $t$ before obtaining any reward is then ,
\begin{eqnarray}
f_{\rm Exponential}\left(t,E\frac{\gamma W c_e}{P_0+\gamma W c_e}\right).
\end{eqnarray}
From this distribution, the expected duration for this game can be expressed as
\begin{eqnarray}
    \bar{t}=\frac{P_0+\gamma Wc_e}{E\gamma W c_e}.
\end{eqnarray}

We can compute the expected reward a miner will earn {\it given that they have won at least one  block in that time period} as
\begin{eqnarray}
    \langle \mathcal{M}^i\rangle\vert_{v^i\ge 1}&=&\frac{\langle\mathcal{M}^i\rangle}{1-f_1(0;p^i,P,E)}\nonumber\\
    &=& \frac{M\frac{\gamma Wc_e}{P_0+\gamma Wc_e}}{1-\sum_{w=0}^\infty\frac{E^we^{-E}}{w!}\left(1-\frac{\gamma Wc_e}{P_0+\gamma Wc_e}\right)^w}.
\end{eqnarray}

Putting together these components in the formula \eqref{eq:agr}, we get our expected growth rate,
The expected growth rate from [2] is then given by
 \begin{eqnarray}\label{eq:main}
&&g(W,\gamma,c_e,c_r,P_0,M,E)\nonumber\\
 &&=\frac{E\gamma Wc_e}{P_0+\gamma Wc_e}\left\{\int_0^{\frac{1-\gamma}{\gamma c_rc_e}}f_{\rm Exponential}\left(t,E\frac{\gamma Wc_e}{P_0+\gamma Wc_e}\right)\right.\nonumber\\
 &&\,\,\,\,\,\,\,\,\,\,\,\,\,\,\,\,\,\,\,\,\,\,\,\,\,\,\,\,\,\,\,\,\times\log\left(\frac{W-t\gamma Wc_rc_e+\langle \mathcal{M}^i\rangle\vert_{v^i\ge 1}}{W}\right)dt.\nonumber\\
&&\,\,\,\,\,\,\,\,\,\,\,\,\,\,\,\,\,\,\,\,\,\,\,\,\,\,\,\,\,\,\,\,\left.+\log(\gamma)\int_{\frac{1-\gamma}{\gamma c_rc_e}}^\infty f_{\rm Exponential}\left(t,E\frac{\gamma Wc_e}{P_0+\gamma Wc_e}\right)\right\}
 \end{eqnarray} 
 where the last term inside the brackets corresponds to the possibilities that the waiting times were longer than the miner could sustain without going bankrupt.

This formula is the main new result of this paper, estimating the growth of a miner's wealth, for a given investment strategy, $\gamma$, and costs of buying vs running equipment, $c_e,c_r$.



Once this expected growth rate is known, the miner can use it to optimize their strategy to maximize their growth.

\subsection{Optimal growth strategy }


Given \eqref{eq:main}, the budget can be chosen such as to maximize the growth for a given {initial} wealth, {$W_0$}. The optimal budget split, $\gamma^*$ can be chosen by maximizing the growth rate,
\begin{eqnarray}\left.\frac{\partial g}{\partial \gamma}\right\vert_{\gamma=\gamma^*}=0,\end{eqnarray}
with the requirement that $\left.\frac{\partial^2 g}{\partial \gamma^2}\right\vert_{\gamma=\gamma^*}<1$

The optimal growth rate is then given by
\begin{eqnarray}g^*(W,c_e,c_r,P_0,M,E)\equiv g(W,\gamma^*,c_e,c_r,P_0,M,E).\end{eqnarray}

Note that mining will only be profitable for large enough miners, mining with insufficient wealth will result in a loss. The minimum miner size, $W_{\rm min}$, is then fixed by
\begin{eqnarray}g^*(W_{\rm min},c_e,c_r,P_0,M,E)=0.\end{eqnarray}
\subsection{Wealth growth for smooth reward}

With smooth rewards, the miner expects to get their corresponding reward after every time period. Define the duration of one period as $\tau$.

This implies a significant rearrangement of the miner optimal budget. The miner only needs to ensure they can run their equipment for $T_{\rm max}=\tau$. The optimal budgeting strategy is then
\begin{eqnarray}\frac{\gamma^*}{1-\gamma^*}=\frac{1}{\tau c_rc_e}.\label{optimalsmoothstrategy}\end{eqnarray} 

The optimal strategy is then to acquire as much consensus power as possible, and leave only a minimal amount of budget to be able to run your equipment for one period. {\it Smooth reward therefore incentivizes the fastest possible growth for all miners}. This can be a powerful insight for protocol designers, which can incorporate elements of smooth reward, in an effort to incentivize fast network growth.

We now compute the expected wealth growth rate using formula \eqref{eq:agr} in the smooth reward scenario, assuming the miner adopts the optimal strategy \eqref{optimalsmoothstrategy}.

The game has only one possible outcome, that with $100\%$ probability each round will last $\tau$, and the reward will be,
\begin{eqnarray}\langle \mathcal{M}^i\rangle=M\frac{\gamma Wc_e}{P_0+\gamma Wc_e}\end{eqnarray}

The optimal growth rate is then
\begin{eqnarray}g^{\rm smooth}(W,c_e,c_r,P_0,M)&=&\frac{1}{\tau}\int_0^{\tau}\log\left(\frac{W-t\gamma^*Wc_ec_r+M\frac{\gamma^*Wc_e}{P_0+\gamma^*Wc_e}}{W}\right)dt
 \nonumber\\
 &=&\frac{1}{\tau}\int_0^{\tau}\log\left(1+\gamma^*\left(M\frac{c_e}{P_0+\gamma^*Wc_e}-tc_ec_r\right)\right)dt.\end{eqnarray}
The dependence in $W$ comes from the denominator containing the miner's contribution to the total network power. In the likely case that $P_0\gg \gamma^*Wc_e$, there is no dependence on $W$, making mining profitable for miners of all sizes.

\subsection{How much should a miner pay to participate in a mining pool?}

We have shown that for small enough miners, mining in the stochastic block reward model will be unprofitable: for small enough wealth, their wealth will decrease over time regardless of strategy. We have also shown that miners of the same kind would still be profitable if they had smooth rewards instead. In less extreme cases, even if miners are large enough to be profitable in both scenarios, profitability is still higher in the smooth rewards scenario. There is therefore quantifiable economic value to the miner in opting for a smooth reward scheme instead of the stochastic case.

Suppose a miner with wealth $W$ chooses to join a mining pool, which will extract a fee in the form of a continuous interest rate $x$, on their rewards. The question we address here is, what is the maximum fee rate $x$ that should be acceptable to this miner, while still extracting positive value from mining?

We first can compute what would happen to the miner's expected wealth over time if they mine optimally, without using the miner pool,
\begin{eqnarray}
    W(t)=W\exp\left[g^*(W,c_e,c_r,P_0,M,E)\cdot t\right].
\end{eqnarray}

We now consider a miner who opted to use the mining pool, which extracts a fraction $x$ of their total rewards. Over time the miner's wealth is expected to grow as
\begin{eqnarray}
    W_{\rm pool}(t)&=&W\exp\left[g^{\rm smooth}(W,c_e,c_r,P_0,M)\cdot t -xt\right]
\end{eqnarray}

The miner should opt to use the mining pool, as long as
\begin{eqnarray}
    W_{\rm pool}(t)>W(t),
\end{eqnarray}
which gives us the condition for the maximum acceptable fee:
\begin{eqnarray}
    x< g^{\rm smooth}(W,c_e,c_r,P_0,M)-g^*(W,c_e,c_r,P_0,M,E).
\end{eqnarray}
As long as the interest rate is below this bound, the miner will be better off joining the mining pool.

Note that the above condition does not guarantee that the miner will be profitable, a small miner could still be unprofitable while participating in the mining pool. If one also requires that beyond the above constraint, the miner should also be profitable, then one also needs to impose $x<g^{\rm smooth}(W,c_e,c_r,P_0,M)$, which is an independent constraint in the case where $g^*(W,c_e,c_r,P_0,M,E)$ is negative.

\section{Conclusions}\label{sec:concl}
In this paper, we investigated the phenomenon of stochastic rewards in a blockchain. Systems such as Bitcoin and Filecoin have stochastic rewards disbursement, meaning that miners may be engaged in the mining process for an extended period of time before receiving a reward. This poses an interesting question for optimal miner behavior, because miners must expend personal resources to continue mining. 

We explore this rigorously in the context of mining pools. Our formulation begins by seeding a miner with a certain amount of starting capital. The miner then decides how to split their capital between a reserve account for covering OpEx, and investing more into CapEx to gain more power and thus increase their probability of winning block rewards. We derive several expressions which dictate optimal miner behavior, where optimality is considered under time-averaged returns rather than ensemble averaged returns.

The research presented herein can be extended in a variety of ways; both from a theoretical and computational perspective. From a theoretical perspective, for example, denoting by $\{Z_n,\ n\in\mathbb{N}\}$ the \textit{cashflow} of a miner at every epoch (a stochastic quantity in the non-smooth case), it is easy to see that $W_n=W_0+\sum_{k=1}^n Z_n$ is also a stochastic process. Furthermore, the mathematical properties of $\{W_n,\ n\in\mathbb{N}\}$ will depend on those of $\{Z_n,\ n\in\mathbb{N}\}$, and it can be shown, under a few mild assumptions that $\{W_n,\ n\in\mathbb{N}\}$ is a Markov process. Based on this alone, one can then use the extensive \textit{machinery} of Markov chains (see, e.g, \cite{meyn2012markov} to derive interesting properties of the wealth process, such as bankruptcy probabilities, expected wealth at $n$-steps, or expected time to bankruptcy, as a function of, e.g., initial choice of strategy or power. From a computational perspective, it would also be interesting to investigate how an initial optimal allocation compares to, e.g., an on-line learned strategy using e.g., reinforcement learning  in the context of agent base simulations \cite{karra2023agent}, or to investigate how the choice of miner strategy affects other parameters in the network.

\bibliography{bib}

\begin{thebibliography}{10}

\bibitem{nakamoto2008bitcoin}
Satoshi Nakamoto.
\newblock {Bitcoin: A peer-to-peer electronic cash system}.
\newblock {\em Decentralized business review}, 2008.

\bibitem{filecoin_whitepaper}
{Protocol Labs}.
\newblock Filecoin: A decentralized storage network.
\newblock Technical report, Protocol Labs, 2017.

\bibitem{wood2014ethereum}
Gavin Wood et~al.
\newblock {Ethereum: A secure decentralised generalised transaction ledger}.
\newblock {\em Ethereum project yellow paper}, 151(2014):1--32, 2014.

\bibitem{stpetersburg_paradox}
Martin Peterson.
\newblock {The St. Petersburg Paradox}.
\newblock In Edward~N. Zalta and Uri Nodelman, editors, {\em The {Stanford}
  Encyclopedia of Philosophy}. Metaphysics Research Lab, Stanford University,
  {F}all 2023 edition, 2023.

\bibitem{peters2011time}
Ole Peters.
\newblock {The time resolution of the St Petersburg paradox}.
\newblock {\em Philosophical Transactions of the Royal Society A: Mathematical,
  Physical and Engineering Sciences}, 369(1956):4913--4931, 2011.

\bibitem{qin2018research}
Rui Qin, Yong Yuan, and Fei-Yue Wang.
\newblock {Research on the selection strategies of blockchain mining pools}.
\newblock {\em IEEE Transactions on Computational Social Systems},
  5(3):748--757, 2018.

\bibitem{chatzigiannis2022diversification}
Panagiotis Chatzigiannis, Foteini Baldimtsi, Igor Griva, and Jiasun Li.
\newblock {Diversification across mining pools: Optimal mining strategies under
  PoW}.
\newblock {\em Journal of Cybersecurity}, 8(1):tyab027, 2022.

\bibitem{liu2018evolutionary}
Xiaojun Liu, Wenbo Wang, Dusit Niyato, Narisa Zhao, and Ping Wang.
\newblock {Evolutionary game for mining pool selection in blockchain networks}.
\newblock {\em IEEE Wireless Communications Letters}, 7(5):760--763, 2018.

\bibitem{liu2017intelligent}
Yi~Liu, Xiayang Chen, Lei Zhang, Chaojing Tang, and Hongyan Kang.
\newblock {An intelligent strategy to gain profit for Bitcoin mining pools}.
\newblock In {\em 2017 10th International Symposium on Computational
  Intelligence and Design (ISCID)}, volume~2, pages 427--430. IEEE, 2017.

\bibitem{wang2019pool}
Yue Wang, Changbing Tang, Feilong Lin, Zhonglong Zheng, and Zhongyu Chen.
\newblock {Pool strategies selection in pow-based blockchain networks:
  Game-theoretic analysis}.
\newblock {\em IEEE Access}, 7:8427--8436, 2019.

\bibitem{qin2020optimal}
Rui Qin, Yong Yuan, and Fei-Yue Wang.
\newblock {Optimal block withholding strategies for blockchain mining pools}.
\newblock {\em IEEE Transactions on Computational Social Systems},
  7(3):709--717, 2020.

\bibitem{schrijvers2017incentive}
Okke Schrijvers, Joseph Bonneau, Dan Boneh, and Tim Roughgarden.
\newblock {Incentive compatibility of bitcoin mining pool reward functions}.
\newblock In {\em Financial Cryptography and Data Security: 20th International
  Conference, FC 2016, Christ Church, Barbados, February 22--26, 2016, Revised
  Selected Papers 20}, pages 477--498. Springer, 2017.

\bibitem{laszka2015bitcoin}
Aron Laszka, Benjamin Johnson, and Jens Grossklags.
\newblock {When Bitcoin mining pools run dry: A game-theoretic analysis of the
  long-term impact of attacks between mining pools}.
\newblock In {\em Financial Cryptography and Data Security: FC 2015
  International Workshops, BITCOIN, WAHC, and Wearable, San Juan, Puerto Rico,
  January 30, 2015, Revised Selected Papers}, pages 63--77. Springer, 2015.

\bibitem{lewenberg2015bitcoin}
Yoad Lewenberg, Yoram Bachrach, Yonatan Sompolinsky, Aviv Zohar, and Jeffrey~S
  Rosenschein.
\newblock {Bitcoin mining pools: A cooperative game theoretic analysis}.
\newblock In {\em Proceedings of the 2015 international conference on
  autonomous agents and multiagent systems}, pages 919--927, 2015.

\bibitem{rosenfeld2011analysis}
Meni Rosenfeld.
\newblock {Analysis of Bitcoin pooled mining reward systems}.
\newblock {\em arXiv preprint arXiv:1112.4980}, 2011.

\bibitem{kroese2013handbook}
Dirk~P Kroese, Thomas Taimre, and Zdravko~I Botev.
\newblock {\em {Handbook of Monte Carlo methods}}.
\newblock John Wiley \& Sons, 2013.

\bibitem{bernoulli22originally}
Daniel Bernoulli.
\newblock {Exposition of a New Theory on the Measurement of Risk. Originally
  published in 1738; translated by Dr. Lousie Sommer.(January 1954).}
\newblock {\em Econometrica}, 22(1):22--36.

\bibitem{meyn2012markov}
Sean~P Meyn and Richard~L Tweedie.
\newblock {\em {Markov chains and stochastic stability}}.
\newblock Springer Science \& Business Media, 2012.

\bibitem{karra2023agent}
Kiran Karra, Tom Mellan, Maria Silva, Juan~P. Madrigal-Cianci, Axel~Cubero
  Cortes, and Zixuan Zhang.
\newblock {An Agent-Based Model Framework for Utility-Based Cryptoeconomies}.
\newblock {\em arXiv preprint arXiv:2307.15200}, 2023.

\end{thebibliography}
\bibliographystyle{unsrt}
\end{document}